\documentstyle[11pt,epsfig,aaspp4]{article}

\def\gtrsim{\mathrel{\hbox{\rlap{\hbox{\lower4pt\hbox{$\sim$}}}\hbox{$>$}}}}

\begin{document}

\title{The Flux Variability of Markarian 501 in Very High Energy
Gamma Rays}

\author{J. Quinn\altaffilmark{1,3}
I. H. Bond,\altaffilmark{2} 
P. J. Boyle,\altaffilmark{3}
S. M. Bradbury,\altaffilmark{2}
A. C. Breslin,\altaffilmark{3}
J. H. Buckley,\altaffilmark{4}
A. M. Burdett,\altaffilmark{1,2}
J. Bussons Gordo,\altaffilmark{3}
D. A. Carter-Lewis,\altaffilmark{5}
M. Catanese,\altaffilmark{5}
M. F. Cawley,\altaffilmark{6}
D. J. Fegan,\altaffilmark{3}
J. P. Finley,\altaffilmark{7}
J. A. Gaidos,\altaffilmark{7}
T. Hall,\altaffilmark{7}
A. M. Hillas,\altaffilmark{2}
F. Krennrich,\altaffilmark{5}
R. C. Lamb,\altaffilmark{8}
R. W. Lessard,\altaffilmark{7}
C. Masterson,\altaffilmark{3}
J. E. McEnery,\altaffilmark{3}
P. Moriarty,\altaffilmark{9}
A. J. Rodgers,\altaffilmark{2}
H. J. Rose,\altaffilmark{2}
F. W. Samuelson,\altaffilmark{5}
G. H. Sembroski,\altaffilmark{7}
R. Srinivasan,\altaffilmark{7}
V. V. Vassiliev,\altaffilmark{1}
and T. C. Weekes\altaffilmark{1}
}

\altaffiltext{1}{ Fred Lawrence Whipple Observatory, Harvard-Smithsonian 
CfA, P.O. Box 97, Amado, AZ 85645-0097, U.S.A.} 
\altaffiltext{2}{Department of Physics, University of Leeds,
Leeds, LS2 9JT, W. Yorkshire, England, U.K.}
\altaffiltext{3}{Department of Experimental Physics, University College, 
Belfield, Dublin 4, Ireland}
\altaffiltext{4}{Deptartment of Physics, Washington University, St. Louis,
MO 63130, U.S.A.}
\altaffiltext{5}{Department of Physics and Astronomy, Iowa State
University, Ames, IA 50011-3160, U.S.A.}
\altaffiltext{6}{Physics Department, National University of Ireland,
Maynooth, County Kildare, Ireland}
\altaffiltext{7}{Department of Physics, Purdue University, West
Lafayette, IN 47907, U.S.A.}
\altaffiltext{8}{Space Radiation Laboratory, California Institute of
Technology, Pasadena, CA 91125, U.S.A.}
\altaffiltext{9}{Department of Physical Sciences, Galway-Mayo Institute of
Technology, Galway, Ireland}

\authoremail{quinn@ferdia.ucd.ie}
\authoraddr{John Quinn, Department of Experimental Physics, University 
College, Belfield, Dublin 4, Ireland}

\begin{abstract}
The BL Lacertae object Markarian 501 was identified as a source of
$\gamma$-ray emission at the Whipple Observatory in March 1995. Here
we present a flux variability analysis on several times-scales of the
233 hour data set accumulated over 213 nights (from March 1995 to July
1998) with the Whipple Observatory 10~m atmospheric \v{C}erenkov
imaging telescope. In 1995, with the exception of a single night, the
flux from Markarian 501 was constant on daily and monthly time-scales and
had an average flux of only 10\% that of the Crab Nebula, making it the
weakest VHE source detected to date. In 1996, the average flux was
approximately twice the 1995 flux and showed significant
month-to-month variability. No significant day-scale variations were
detected. The average $\gamma$-ray flux above $\sim$350~GeV in the
1997 observing season rose to 1.4 times that of the Crab Nebula -- 14
times the 1995 discovery level -- allowing a search for variability on
time-scales shorter than one day. Significant hour-scale variability
was present in the 1997 data, with the shortest, observed on MJD
50607, having a doubling time of $\sim$2 hours.  In 1998 the average
emission level decreased considerably from that of 1997 (to $\sim$20\%
of the Crab Nebula flux) but two significant flaring events were
observed. Thus, the emission from Markarian 501 shows large amplitude
and rapid flux variability at very high energies as does Markarian
421. It also shows large mean flux level variations on year-to-year
time-scales, behaviour which has not been seen from Markarian 421 so far.
 
\end{abstract}

\keywords{BL Lacertae objects: individual (Markarian 501) --- gamma-rays:
observations}

\section{Introduction}
Three active galactic nuclei (AGN) have been discovered to be very
high energy (VHE, E$\gtrsim$300 GeV) $\gamma$-ray sources by the
Whipple Observatory $\gamma$-ray collaboration: Markarian 421
(Mrk~421) (\cite{Punch92}), Markarian 501 (Mrk~501)(\cite{Quinn96})
and 1ES 2344+514 (\cite{Catanese98}).  These are the three closest BL
Lacertae objects (BL Lacs), with redshifts in the range 0.0308 -
0.044, and are among the brightest at X-ray energies. They are
all classified as X-ray selected BL Lacs (XBLs) as their
synchrotron spectra extend into the X-ray range. A fourth BL Lac,
PKS2155-304, has been detected in VHE $\gamma$-rays by the University
of Durham group (\cite{Chadwick99}).

The Energetic Gamma-Ray Experiment Telescope (EGRET) on board the
Compton Gamma-Ray Observatory has detected at least 51 AGN at energies
$>$100~MeV (\cite{Thompson95}; \cite{Mukherjee97}). They are
all members of the blazar class of AGN, which include flat spectrum
radio quasars and BL Lacs. Of the EGRET-detected blazars, 14 are
BL Lacs with 12 Radio Selected BL Lacs (RBLs) and only
two XBLs. Mrk~421 is the only VHE source in this catalogue and
it is among the weakest. However, Mrk~501 has recently been
detected at the 4$\sigma$ level with EGRET (\cite{Kataoka99}). One of
the most striking characteristics of the EGRET-detected blazars
is variability; 42 of the 51 AGN exhibit variability
(\cite{Mukherjee97}).  Variability time-scales as short as 4 hours have been
observed (\cite{Mattox97}).

VHE observations of Mrk~421 and Mrk~501 have also revealed extreme
variability (e.g. \cite{Gaidos96}; \cite{Quinn96}).  The VHE flux has
been measured to vary by nearly a factor of 100 in Mrk 421
(\cite{McEnery99}) and, as we show in the following sections, Mrk 501
has been measured with fluxes ranging from 0.1 to 5 times the Crab
Nebula flux with the Whipple Observatory atmospheric \v{C}erenkov
telescope.  The large collection area ($\sim 3.5\times 10^{5}$~m$^2$)
of the Whipple Observatory telescope permits sensitive studies of
variability on time-scales inaccessible to space-based
telescopes. Indeed, the shortest observed variability of any blazar at
any $\gamma$-ray energy, a 30 minute duration flare observed from
Mrk~421 (\cite{Gaidos96}), was measured with this telescope. Both
Mrk~421 and Mrk~501 have been closely monitored by the Whipple
Collaboration since their discovery, with an $\sim$0.5 hr
exposure/night being sufficient for detection of flaring
activity.  Prior to 1997, the $\gamma$-ray emission from Mrk~421 was
generally observed to have a higher mean flux (\cite{Schubnell96}) and
to have been more frequently variable (\cite{Buckley96}) than that of
Mrk~501.

In Spring 1997, the Whipple Collaboration observed Mrk~501 to be in an
unprecedented high emission state at VHE energies, as subsequently
confirmed by several independent \v{C}erenkov imaging groups
(\cite{Protheroe98}).  A public Compton Gamma-Ray Observatory Target
of Opportunity was initiated in response to a request by the Whipple
Collaboration in 1997.  Evidence of correlated variability in data
from the Whipple Observatory $\gamma$-ray telescope, the Oriented
Scintillation Spectrometer Experiment (OSSE) and the All-Sky Monitor
(ASM) of the {\sl Rossi X-ray Timing Explorer} is presented in
Catanese et al.\ (1997). During this observing campaign the energy output of
Mrk 501 in VHE $\gamma$-rays was comparable to that in the 2-100 keV
range but the variability amplitude was larger. The correlations seen
may imply some relativistic beaming of the emission, given that the
spectrum extends to $\gtrsim$7 TeV (\cite{Samuelson98}). There was
also some indication that the optical U-band flux was higher on average
in the month of peak $\gamma$-ray activity. 

Here, based on 4 years of data, we present a study of the variability
of the $\gamma$-ray flux above $\sim$350~GeV from Mrk 501. Details of
the observations are given in \S\ref{section:observations} and the
analysis methodology, including $\gamma$-ray selection criteria and
the test for variability, is described in
\S\ref{section:analysis}. The results of the analysis are presented in
\S\ref{section:results} and their implications briefly discussed in
\S\ref{section:discussion}.

\section{Source Observations}
\label{section:observations}
Observations were made with the Whipple Observatory 
10~m atmospheric \v{C}erenkov imaging telescope (\cite{Cawley90}),
located on Mt. Hopkins in southern Arizona.  The 10~m reflector
images the \v{C}erenkov radiation from cosmic-ray and $\gamma$-ray
initiated air-showers onto a high resolution camera  mounted in the
focal plane.  Subsequent off-line analysis of the images (described
below) facilitates the selection of candidate $\gamma$-ray events. 

The high resolution camera utilises fast
photomultiplier tubes (PMTs) arranged in a hexagonal array, with
inter-tube spacing of $0\fdg25$. During 1995 and 1996 the camera
consisted of 109 PMTs with a resulting field of view (FOV) of $\sim
2\fdg8$. An event was recorded when any two of the inner 91 PMTs
registered a signal $\gtrsim$40 photoelectrons within an effective
resolving time of 15~ns. For the 1997 observations a further 42
PMTs were added, resulting in a FOV of $\sim 3\fdg4$.
The trigger condition remained the same as for
the 109 PMT camera. In Summer 1997 a new camera, containing 331
pixels, was installed.  This camera has a FOV of $\sim 4\fdg8$. The
trigger condition for this enlarged camera required that any two of
the 331 pixels produce a signal $\gtrsim$40 photo-electrons within an
effective resolving time of 8~ns. The telescope was also triggered
artificially, after every 24 events for the 109/151 pixel cameras and
once every second for the 331 pixel camera, to determine the
background sky-brightness level in each PMT. Light-cones, which
minimise the dead-space between PMTs and reduce the albedo effect,
were used on the 109 and 151 pixel cameras, but were not yet 
installed on the 331 pixel camera.

In general, two modes of observation are used: \textsc{on/off} and
\textsc{tracking}. With the \textsc{on/off} mode the source is 
tracked continuously for 28 minutes and then, to estimate the
background, a region offset in right ascension (RA) by 30 minutes
(allowing 2 minutes slew time) is tracked. This has the
disadvantage that an equivalent amount of observation time is spent
looking away from the source.  Alternatively, the \textsc{tracking}
mode, where the background is estimated from the \textsc{on} source
run itself (see \S\ref{section:trkanalysis}), can be used. In this
case an \textsc{off} source run is not required for each \textsc{on}
source run, allowing continuous monitoring of an object. However,
\textsc{off} runs are still needed to determine the response of the
telescope to background events.

Observations are typically made when the source zenith angle is less
than 35$^\circ$ and are referred to as small zenith angle (SZA)
observations. Observations at large zenith angles (LZA,
typically 55$^\circ$ to 70$^\circ$) may also be made. Increasing the
zenith angle has the effect of increasing the energy threshold,
but has the benefit of increasing
the collection area. Thus, it is an excellent method for increasing 
photon statistics to facilitate the determination of the energy spectrum
at higher energies. For a detailed description of the LZA technique
see Krennrich et al.\ (1997).

Since its discovery as a $\gamma$-ray source in 1995 (\cite{Quinn96}),
the VHE $\gamma$-ray emission from Mrk~501 has been monitored
intensively with the Whipple Observatory 10~m telescope. Only SZA 
observations taken under good sky conditions have been considered in the 
analysis for variability presented here. Our selection includes data from 56, 50, 
55 and 49 nights of observation in the Spring -- Summer periods
of 1995, 1996, 1997 and 1998 respectively.
Table~\ref{table:observations} summarises the resulting database.

\placetable{table:observations}

\section{Data Analysis}
\label{section:analysis}
\subsection{$\gamma$-ray Selection}
The vast majority of events detected by \v{C}erenkov telescopes are
cosmic rays.  Candidate $\gamma$-ray events are selected on the basis
of the shape and orientation of the \v{C}erenkov images: $\gamma$-ray
images are typically more compact and elliptical than background
hadronic images and tend to have their major axes aligned with the
source location in the FOV. Background cosmic ray events, on the
other hand, have random orientations.

Each image is first subjected to a cleaning procedure (\cite{Fegan97})
which suppresses pixels which are dominated by light from fluctuations
of the night-sky background. A moment-fitting routine is then used to
calculate various image parameters.  The shape of each image is
characterised by the parameters \emph{length} and \emph{width} and the
orientation by $\alpha$, the angle between the major axis of the image
and the line joining the source location in the FOV to the centroid of
the image. In addition, for the data taken in 1998, an
\emph{asymmetry} cut has been included.  $\gamma$-ray images have a
cometary shape with a tail which points away from the source location
in the FOV (\cite {Buckley98}) and thus their intensity profiles have
positive asymmetry. The larger FOV of the camera used in 1998 allows
this parameter to be accurately determined, something which was not
possible with the smaller FOV cameras.

Before the application of shape and orientation cuts a 
\emph{software trigger} cut is also applied to eliminate events close to
threshold, some of which are induced by noise fluctuations. The
software trigger involves cuts on the image \emph{size} (i.e. the
total number of photoelectrons recorded), the counts in each of
the brightest two tubes (\emph{max1, max2}), and a requirement that at least
three tubes above a low noise threshold (2.25$\sigma$, where $\sigma$
is the RMS sky-noise in a PMT, as determined from the artificially
triggered events) be neighbours (\emph{NBR3}). A \emph{distance} cut
is applied to eliminate images which are too close to the center of
the camera and will have poor $\alpha$ reconstruction and also those
events which have occurred too close to the edge of the FOV and may be
truncated.

Due to the continuous evolution of the high resolution camera, and
changes such as deterioration of mirror reflectivity through
weathering and the presence or absence of light-cones, the optimum
data analysis cuts differ for each year. The cuts used for
a given telescope configuration are optimised on an independent
data set, usually data taken on the Crab Nebula or, in the case of the
1998 data, Mrk~421.  Table \ref{table:cuts} lists the cuts used
for each years' analyses.

\placetable{table:cuts}

\subsection{$\gamma$-ray Rate and Flux Calculation}
\label{section:trkanalysis}
There are slightly different analysis methods for the \textsc{on/off}
and \textsc{tracking} observation modes. For the \textsc{on/off}
observations the background is estimated from the \textsc{off} source
run, which is assumed to be on a sky region which does not include a
$\gamma$-ray source.  This analysis mode has been discussed at length
elsewhere (e.g.  \cite{Kerrick95}; \cite{Catanese98}).

For \textsc{tracking} observations the background is estimated
from the \textsc{on} source run itself. All of the $\gamma$-ray
selection criteria apart from orientation ($\alpha$) are applied to
the data.  The background is then estimated from events which are not
oriented towards the source.  In this analysis, background events with
values of $\alpha$ between 20$^\circ$ and 65$^\circ$ are used. Images
having values of $\alpha$ between 65$^\circ$ and 90$^\circ$ are
discarded because of possible systematic effects due to truncation at
the camera's edge. Once the number of events with orientations in
the 20$^\circ$ to 65$^\circ$ range is known then the expected number
of background events in the \emph{signal} domain ($\alpha$ of
0$^\circ$ to 15$^\circ$, or to 10$^\circ$ for data taken in 1998) can
be estimated. \textsc{Off} source data recorded for this source 
and others can be combined to calculate a ratio,
$r\,\pm\,\Delta r$, of the number of events in the signal region to
those in the 20$^\circ$ to 65$^\circ$ region in the absence of a
source. In this case, the significance of a $\gamma$-ray excess, $S$,
is given by:

\begin{equation}
	S = \frac{N_{on} - r\, N_{off}}
             {\sqrt{N_{on} + r^{2} N_{off} + (\Delta r)^{2}N_{off}^{2}}}
\end{equation}

\noindent and the $\gamma$-ray rate ($R\,\pm\,\Delta R$)
is calculated from:

\begin{equation}
	R\pm\Delta R = \frac{N_{on}- r\, N_{off}}{t} \pm 
	                      \frac{\sqrt{N_{on} + r^{2} N_{off} + 
                                     (\Delta r)^{2}N_{off}^{2}}}{t} 
\end{equation}

\noindent where $N_{on}$ is the number of counts in the $\gamma$-ray domain
($\alpha\ < \ 10^\circ$ or $15^\circ$), $N_{off}$ is the number of
counts in the 20$^\circ$ to 65$^\circ$ $\alpha$ range and $t$ is the
duration of the observation. 

The inclusion of the statistical error on the \textsc{tracking} ratio
effectively limits the amount (duration) of \textsc{tracking} data
which can be usefully analysed for an excess. Once the duration of
the \textsc{tracking} data exceeds that of the \textsc{off} source data
used in the calculation of the tracking ratio the error on the
tracking ratio starts to dominate and to limit the significance of a
detection.

For the purpose of investigating possible differences in results
produced by the \textsc{on/off} and \textsc{tracking} analyses, 
the Crab Nebula data were analysed with both methods (\cite{Quinn97}).
The results demonstrated that the $\gamma$-ray rates derived using 
both methods were consistent and stable; the tracking analysis did
not introduce any apparent variability in the rate.  For the analysis
presented in this paper, all of the \textsc{on} source data were
combined with the \textsc{tracking} data and the resulting database
analysed using the \textsc{tracking} analysis.

The data presented here were taken with different telescope
configurations having different sensitivities. It is therefore
necessary to normalize when comparing the different data sets. To a
first approximation, this can be achieved by converting the rates to
fractions of the $\gamma$-ray rate from the Crab Nebula taken with the
same telescope configuration and analysed with the same cuts. The Crab
Nebula is believed to be a steady source of VHE $\gamma$-rays, as has
been observed by the Whipple Collaboration over the past decade
(\cite{Cawley99}).  To convert a given $\gamma$-ray rate to an
integral flux, the rate as a fraction of the Crab Nebula flux was
multiplied by $(1.05 \pm 0.24)\times 10^{-10}\ $cm$^{-2}$s$^{-1}$,
which represents the integral Crab Nebula flux above 350~GeV
(\cite{Hillas98}), the threshold of the analysis presented here.

\subsection{Test for Variability}
To search for temporal variability in the $\gamma$-ray flux 
we apply a $\chi^2$ test for a constant rate.
The $\chi^2$ sum is converted into a probability ($P_{\chi^2}$) that
the emission is constant about the mean using the incomplete gamma
function \emph{gammq}(a,x) (\cite{Press_recipes}) 
\begin{equation}
P_{\chi^2}=gammq \left(\frac{N-1}{2},\frac{\chi^2}{2}\right)
\end{equation}
where N-1 is the number of degrees of freedom. The number of trials is
taken into account by calculating the probability ($P_{trials}$) of
$P_{\chi^2}$ occurring in N trials from:
\begin{equation}
P_{trials}=1-(1-P_{\chi^2})^{N}.
\end{equation}

\noindent This method is used to test whether the distribution of 
measured $\gamma$-ray rates is consistent with statistical 
fluctuations about the mean for a range of time-scales. 

\section{Results}
\label{section:results}

The data have been analysed using the \textsc{tracking} analysis described above.
This differs from that applied to the 1995 data by Quinn et al. (1996), in that
a statistical error on the \textsc{tracking} ratio is now included and that a 
10\% systematic error is no longer added to \textsc{tracking} results
as a careful study showed that the results of the \textsc{on/off} and
\textsc{tracking} analysis methods are in close agreement
(\cite{Quinn97}). In fact, these changes tend to cancel each other.

For the 1995 data set as a whole we obtain a $\gamma$-ray rate of approximately 
10\% of that of the Crab Nebula ($0.18\,\pm\,0.02\ $min$^{-1}$, giving a  
9.1$\sigma$ excess). This rose to approximately 20\% of the Crab Nebula rate 
for the following season (analysis of 1996 data reveals an excess of 11.1$\sigma$ and 
a $\gamma$-ray rate of $0.26\,\pm\,0.02\ $min$^{-1}$), indicating 
that the average emission level had doubled since the previous year. 

The monthly and nightly average rates, in fractions of the Crab Nebula rate, 
over all four years of Mrk 501 observation are shown in Figure \ref{fig-m5all}.
The rate in 1995 appears to have been constant with the exception
of one night, MJD 49920, when the it was approximately 4.6$\sigma$
above the average. A $\chi^2$ test gives a chance probability of
$1.2\times 10^{-3}$ (after accounting for trials) that the daily
averages were constant during that month. The $\chi^2$
probability that the daily averages are constant over the entire 5
months of observation is 0.17 (after trials). The probability that the
emission is constant when averaged on monthly time-scales is 0.06
(after trials). Conversely, in the 1996 data set there were no obvious
flaring episodes, but when averaged on time-scales of a month there is
significant variability. The $\chi^2$ probability that the emission is
constant for the monthly averages is $3.8\,\times\,10^{-6}$, after
accounting for trials. When each month is examined for variability
with the rates averaged on time-scales of a day no significant
variability is found. However, the $\chi^2$ probabilities are smaller
than for a similar analysis of the 1995 data, suggestive of increased
day-scale flickering in the $\gamma$-ray emission.
The results of the $\chi^2$ test for variability of
the monthly and daily averages are shown in Tables
\ref{table:monthly_var} and \ref{table:daily_var} respectively.

\placefigure{fig-all}

Observations of Mrk~501 in 1997 began in January using the LZA 
technique (\cite{Krennrich98}). The initial observations 
suggested that the flux was much higher than in
previous years. Conventional SZA observations  
commenced as soon as possible in February.
Other ground-based $\gamma$-ray experiments were notified and
high emission levels were verified. In March
a joint IAU circular by the CAT, HEGRA and Whipple groups (\cite{Breslin97}) 
announced preliminary results.
The Whipple observations continued through June and included
observations made with the LZA technique. The results of the LZA data
are presented elsewhere (\cite{Krennrich98}).  

Our SZA observations revealed that the VHE $\gamma$-ray emission was very 
strong throughout Spring -- Summer 1997. Significant variability was observed 
in the monthly averages. The emission appeared to increase steadily from 
February through May and then to level off in June.  The daily rates also 
exhibit dramatic variability. Day-to-day changes in the flux by factors $>$4
were observed and on eight occasions the flux more than doubled
between consecutive nights. On four occasions there were equally
rapid decays in the rate. The average
flux level for the season is 1.4 times that of the Crab Nebula - an
increase by a factor of 14 from the level in 1995. The peak rate,
observed on MJD 50554, is 3.7 times the Crab rate. The average rates
for the 1997 data were calculated from only one run (the first with
elevation above 55$^\circ$) per night. This was done to remove a bias
in the observing strategy whereby observations of Mrk~501 on a given
night continued only if the source was in a very active state.

\begin{figure} 
\centerline{\epsfig{file=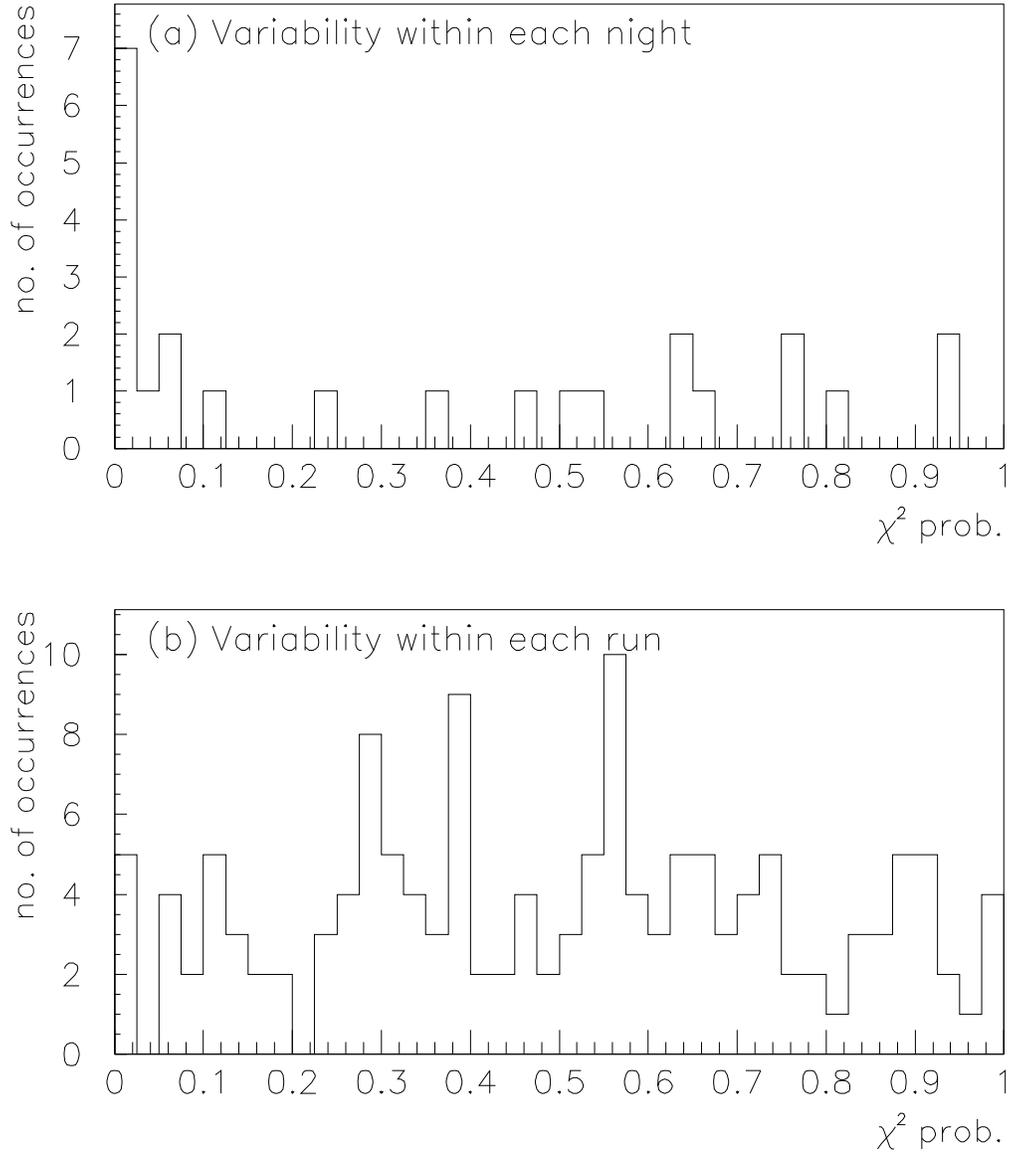,height=6in,angle=0.}} 
\caption{Distribution of $\chi^2$ probabilities for constant emission
for (a) each of 24 nights (28 min. bins) and (b) each of 144 runs (9
min. bins) for data taken in 1997. There is a significant excess of
small probabilities in plot (a), indicating hour-scale variability.
\label{fig-m597-csqdist}}
\end{figure}

\begin{figure} 
\centerline{\epsfig{file=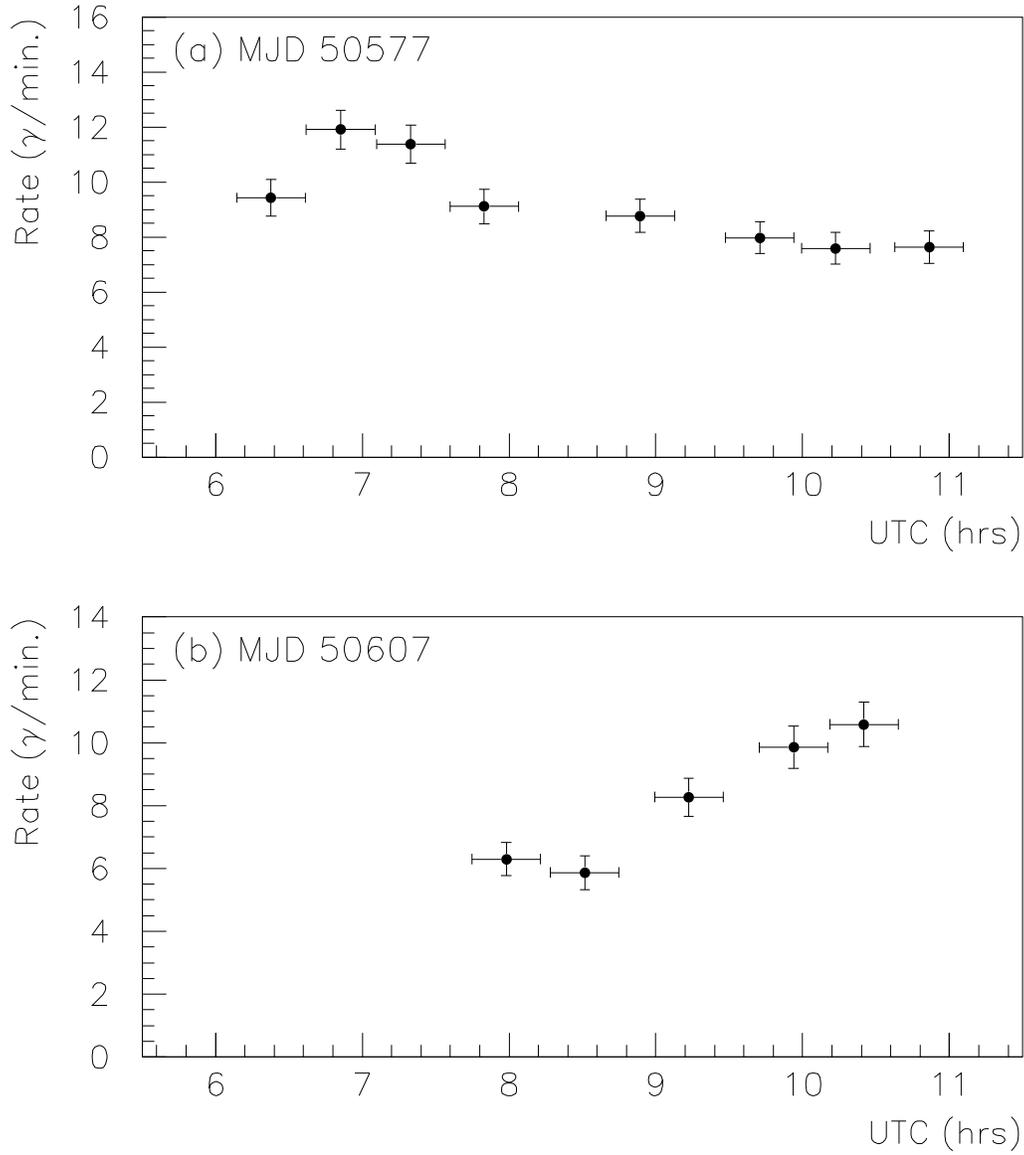,height=6in,angle=0.}} 
\caption{$\gamma$-ray rates for the two nights in 1997 which show significant
variability on a time-scale of hours. The probability (after accounting
for trials) for constant emission during May 9 (a) is
$5.2\times10^{-6}$ while for June 8 (b) it is $5.8\times10^{-8}$.  The
flux on the latter night has a doubling time of $\sim$2 hours.
\label{fig-hsvar-strong}}
\end{figure}

The strong signal-to-background level in the 1997 data allowed a
search to be made for variability on time-scales shorter than one
day. For this test, data from each night on which there were three or
more runs (approx. 1.5 hours, see \S\ref{section:observations}) were
analysed to test for run-to-run variability. A total of 24 nights
satisfied this criterion. The $\chi^2$ probability that the emission
was constant was calculated for each of the nights; the distribution
of these probabilities is shown in Figure
\ref{fig-m597-csqdist}a. For a statistically-variable source this
distribution should be flat but there is an excess of nights with
small probabilities.  Assuming that the variations are purely
statistical, the probability of getting seven nights in this first bin
(width = 0.025) out of 24 trials is 1.45$\times$10$^{-6}$.  Of these
seven nights, two show statistically significant variations within
themselves. The probability for constant emission on MJD 50577 is
$5.2\,\times\,10^{-6}$ while for MJD 50607 the probability is
$5.8\,\times\,10^{-8}$ (after accounting for trials). The flux on MJD
50607 has a doubling time of $\sim$2 hours. We thus identify these two
nights as exhibiting significant hour-scale variability while the five
other nights exhibit marginal variability. Figure
\ref{fig-hsvar-strong} shows the $\gamma$-ray 
rates for the two nights with significant variability.

\placefigure{fig-m597-csqdist}
\placefigure{fig-hsvar-strong}

A search of the 1997 data for variability on time-scales of less than half 
an hour has also been performed. For this test each of the 28 minute runs, 
144 in total, was divided into three equal length intervals.
Each triplet was then analysed for variability. No significant
variations were found and the distribution of $\chi^2$ probabilities
(Figure \ref{fig-m597-csqdist}b) does not indicate any excess of low
probabilities. Hence, we see no evidence for significant sub-hour 
scale variability.

The average $\gamma$-ray rate for the 1998 data set was 0.42$\pm$0.04~min$^{-1}$,
approximately 20\% of the rate obtained from the Crab Nebula
i.e. on average the emission was much lower than in 1997. In fact, the 
average rates for March, April and May are comparable to the level of the 
initial detection in 1995. There were however two significant flaring events. 
The first occurred in early March where an apparent rise and decay were
observed.  Unfortunately this flare is poorly sampled due to bad
weather.  The flux was observed to be relatively high ($\sim$1.3 times
the Crab Nebula flux) on MJD 50876 and on the following
night a flux of $\sim$5.0 times that of the Crab Nebula was recorded.
This is the largest flux detected to date from Mrk~501 by the
Whipple Observatory 10~m telescope.  For the next observation
on MJD 50880 the measured flux was 
still relatively high ($\sim$1.3 times the level of the Crab
Nebula flux).  A second flare occurred in June. The flux increased on two
consecutive nights, peaking at $\sim$1.1 times the Crab
Nebula flux on MJD 50991, then decayed on a similar
time-scale.  The average flux level immediately to either side of
this flare was below the sensitivity of the telescope.

\section{Discussion}
\label{section:discussion}

\begin{figure} 
\centerline{\epsfig{file=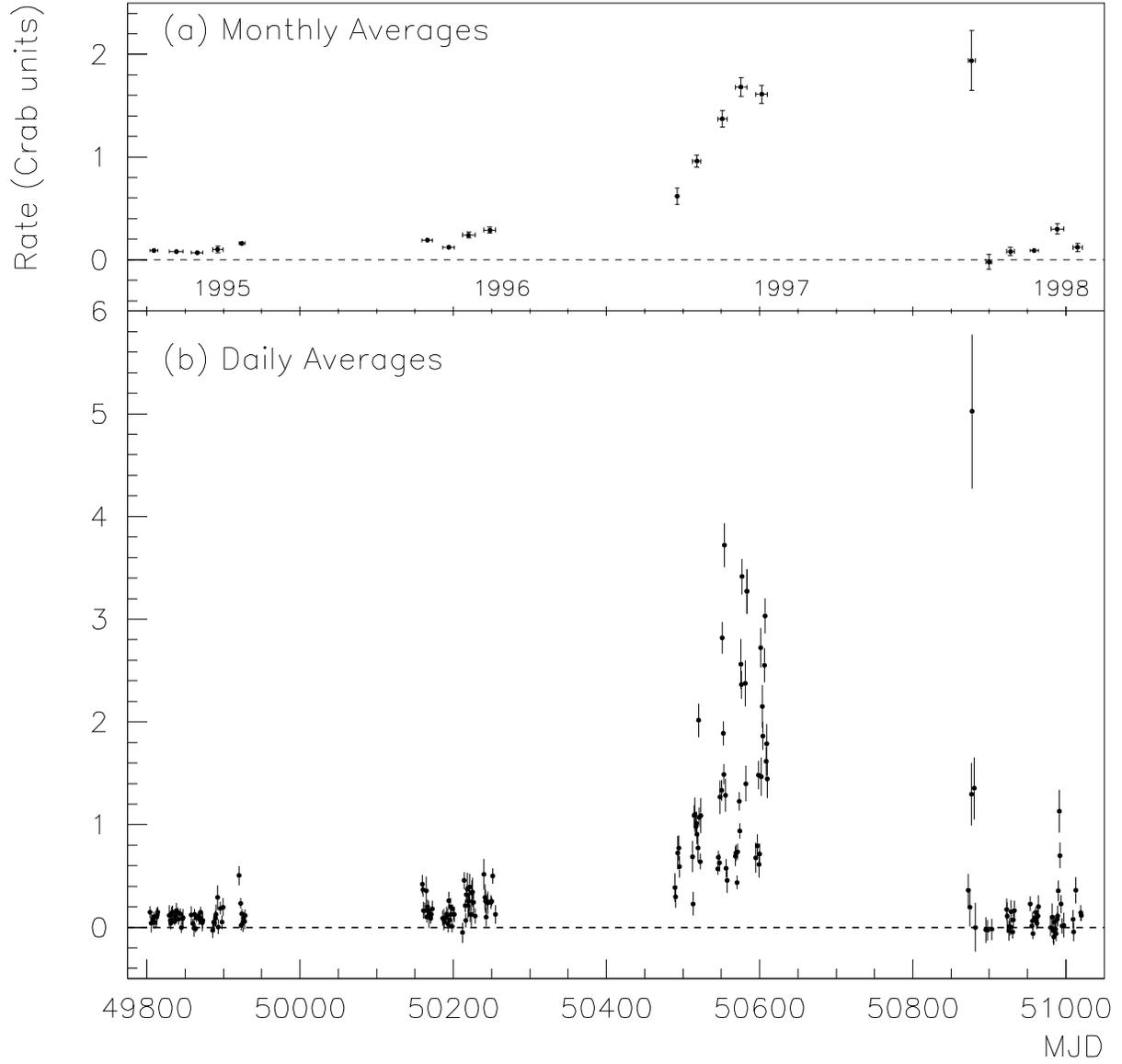,height=6in,angle=0.}} 
\caption{Average $\gamma$-ray rates on (a) monthly and (b) daily
time-scales for Mrk~501 between  1995 and 1998.
\label{fig-m5all}}
\end{figure}

We have demonstrated that rapid variability, a common characteristic
of blazars at all observed energies, is also present in the VHE
$\gamma$-ray emission from Mrk 501.  Our 4 year data set spans a
remarkable change in the flux level from Mrk 501: the average 
yearly emission
level exhibited a fourteen-fold increase between 1995 and 1997 and
the average daily  flux varied by a factor of $\sim$50 (see
Figure \ref{fig-m5all}). In 1997 large amplitude day-scale flares
occurred frequently and were usually followed by equally rapid
decays. Day-scale changes in flux by factors as large as 4.7 were
observed. Episodes of significant hour-scale variability were
detected, with one having a doubling time of 2 hours. In addition,
there is evidence of consistent hour-scale variability which is not
resolved in individual episodes.  A variable flux
from Mrk 501 was reported by at least four other atmospheric
\v{C}erenkov observatories in 1997 (see, e.g. \cite{Protheroe98}).

The data presented here suggest that Mrk~501 was more variable when
the flux level was higher. However, this effect could also be due to
the sensitivity of the telescope. At low flux levels it
takes longer to accumulate a significant signal,
so the search for short-term variability may be limited by poor
statistics. To address this issue we performed a test to see if
the day-scale variability observed in 1997 would have been detected in
1996 and/or 1995 and if the month-scale variability seen in 1996 and
1997 would have been detected in 1995. We calculated
the percentage deviations about the mean level from a period where
significant variability was observed and then, using the mean signal
and background level from another period, calculated the signal (and
statistical error) which would have been observed given the same
percentage deviations about that mean. The results are that 
neither the 1997 month-scale nor day-scale variability would have been
detected at a significant level if present in 1995. However, the 1996
month-scale variability would have been significant if
present in 1995 (chance probability of $\sim 10^{-6}$), while the 1997
degree of day-scale variability, if present in 1996, would have been
detectable (chance probability $\sim 10^{-7}$).
Thus we conclude that there was a change in the flaring
characteristics, in addition to the change in the mean flux level
between the different observing seasons.

The VHE $\gamma$-ray emission from Mrk~501 exhibits rapid variability
similar to that of the emission from Mrk~421. A major
difference between the two objects is that the emission from Mrk~501
seems to have a base-level, which changes on monthly and yearly
time-scales, whearas the VHE $\gamma$-ray emission from Mrk~421
has been described as consisting of a series of rapid flares
with no underlying baseline (\cite{Buckley96}).  The
variability of the VHE $\gamma$-ray emission of Mrk~501 in 1995 and
1996 was similar to that at other wavelengths, with small
amplitude, slow variations being more common than fast, large
amplitude flares.

The increase in the VHE $\gamma$-ray power and variability in 1997
was accompanied by an increase in the hard X-ray power and an
extension of the synchrotron spectrum to at least 100~keV
(\cite{Catanese97}).  This is consistent with an
inverse-Compton mechanism for producing VHE $\gamma$-rays.  However,
our observations of Mrk~501 to date cannot discriminate between the
inverse-Compton (electron) models and those where the dominant
particles producing $\gamma$-rays in the jet are protons. More densely
sampled light curves, covering as broad a wave-band as possible, are
needed to provide more insight into the mechanisms responsible for the
VHE radiation from BL Lac objects. 
Future, more sensitive $\gamma$-ray observations of AGN with proposed
detectors such as GLAST, HESS, MAGIC and VERITAS will allow the
structure of flares on shorter time-scales to be determined.

\vspace{0.5cm}
\acknowledgments

We acknowledge the technical assistance of K. Harris and E. Roache. This
research is supported by grants from the US Department of Energy and by
NASA, by PPARC in the UK and by Forbairt in Ireland.

\begin{deluxetable}{lrrrr}
\tablecaption{Summary of observations (combined hours of 
\textsc{on} and \textsc{tracking} data)}
\tablehead{
\colhead{} & \colhead{1995} & \colhead{1996} &
               \colhead{1997}& \colhead{1998} 
}
\tablewidth{0pt}
\startdata
Feb              & \nodata  & \nodata & 3.3	&  0.9  \nl
Mar              &  3.7     & 18.0    & 12.4    &  5.5  \nl
Apr              & 17.7     & 15.5    & 20.7    &  4.6  \nl
May              & 20.3     & 10.3    & 21.2    & 20.5  \nl
Jun              & 11.5     &  8.8    &  9.7    & 11.6  \nl
Jul              &  8.3     & \nodata & \nodata &  8.7  \nl 
\tableline
Total            & 61.5     & 52.6    & 67.3    & 51.8  \nl
\enddata  
\label{table:observations}
\end{deluxetable}

\begin{deluxetable}{cc}
\tablecaption{$\gamma$-ray selection critera}
\tablehead{
\colhead{Data set} & \colhead{Cuts}
}
\tablewidth{45ex}
\startdata
     & 0.073$^\circ$ $\leq$ {\em width} $\leq$ 0.15$^\circ$   \nl
     & 0.16$^\circ$ $\leq$ {\em length} $\leq$ 0.30$^\circ$   \nl
1995 & 0.51$^\circ$ $\leq$ {\em distance} $\leq$ 1.0$^\circ$  \nl
 and & $\alpha \leq 15^\circ$                                 \nl
1996 & \em{size} $>$ 400 d.c.                                 \nl
     & \em{max1} $>$ 100 d.c.                                 \nl
     & \em{max2} $>$ 80 d.c.                                  \nl
\tableline
     & 0.073$^\circ$ $\leq$ {\em width} $\leq$ 0.16$^\circ$   \nl
     & 0.16$^\circ$ $\leq$ {\em length} $\leq$ 0.33$^\circ$   \nl
     & 0.51$^\circ$ $\leq$ {\em distance} $\leq$ 1.17$^\circ$ \nl
1997 & $\alpha \leq 15^\circ$                                 \nl
     & \em{size} $>$ 0 d.c.                                   \nl
     & \em{max1} $>$ 95 d.c.                                  \nl
     & \em{max2} $>$ 45 d.c.                                  \nl
\tableline
     & 0.073$^\circ$ $\leq$ {\em width} $\leq$ 0.16$^\circ$   \nl
     & 0.16$^\circ$ $\leq$ {\em length} $\leq$ 0.44$^\circ$   \nl
     & 0.51$^\circ$ $\leq$ {\em distance} $\leq$ 1.25$^\circ$ \nl
1998 & $\alpha \leq 10^\circ$                                 \nl
     & \em{size} $>$ 0 d.c.                                   \nl
     & \em{max1} $>$ 75 d.c.                                  \nl
     & \em{max2} $>$ 65 d.c.                                  \nl
     & \em{asymmetry} $>$ 0                                        \nl
\enddata  
\label{table:cuts}
\end{deluxetable}

\begin{deluxetable}{lcrr}
\tablecaption{Results of analysis for variability of the monthly averages}
\tablewidth{0pt}
\tablehead{
\multicolumn{1}{l}{Year} 
&\multicolumn{1}{c}{Number of  Months} 
&\multicolumn{1}{c}{Rate \tablenotemark{a} }
&\multicolumn{1}{c}{P$_{trials}$ \tablenotemark{b}} 
}
\startdata
1995      &  5 & 0.10$\pm$0.01 & 0.05  \nl
1996      &  4 & 0.20$\pm$0.02 & 3.6$\times$10$^{-5}$   \nl
1997      &  5 & 1.39$\pm$0.07 & 8.8$\times$10$^{-25}$   \nl
1998      &  6 & 0.32$\pm$0.04 &  8.5$\times$10$^{-11}$ \nl
\tablenotetext{a}{Average rate over the observation period
expressed as a fraction of the rate from the Crab Nebula.}
\tablenotetext{b}{$\chi^2$ probability for constant emission after 
accounting for trials.}
\enddata  
\label{table:monthly_var}
\end{deluxetable}

\begin{deluxetable}{lcrr}
\tablecaption{Results of analysis for variability of the daily averages}
\tablewidth{0pt}
\tablehead{
\multicolumn{4}{c}{1995} 
}
\startdata
\multicolumn{1}{l}{Obs. Period} 
&\multicolumn{1}{c}{Number of Nights} 
&\multicolumn{1}{c}{Rate \tablenotemark{a} }
&\multicolumn{1}{c}{P$_{trials}$ \tablenotemark{b}} \nl 
\tableline
Mar 28 - Apr 07 & 09 & 0.09$\pm$0.02 & 1.00  \nl
Apr 22 - May 10 & 15 & 0.08$\pm$0.02 & 1.00  \nl
May 21 - Jun 05 & 14 & 0.07$\pm$0.02 & 1.00  \nl
Jun 18 - Jul 01 & 10 & 0.10$\pm$0.03 & 0.99  \nl
Jul 22 - Jul 30 & 08 & 0.16$\pm$0.02 & 1.8$\times$10$^{-3}$ \nl
Total           & 56 & 0.10$\pm$0.01 & 0.22   \nl
\tableline
\tablevspace{1ex}
\multicolumn{4}{c}{1996}  \nl
\tablevspace{1ex}
\tableline
Mar 17 - Mar 30 & 13 & 0.19$\pm$0.02 & 0.33  \nl
Apr 13 - Apr 28 & 15 & 0.12$\pm$0.02 & 1.00  \nl
May 09 - May 25 & 15 & 0.24$\pm$0.03 & 0.18  \nl
Jun 06 - Jun 21 & 10 & 0.29$\pm$0.03 & 0.46  \nl
Total      & 53 & 0.26$\pm$0.02 & 3.8$\times$10$^{-6}$ \nl

\tableline
\tablevspace{1ex}
\multicolumn{4}{c}{1997}  \nl
\tablevspace{1ex}
\tableline
Feb 10 - Feb 16 & 05 & 0.62$\pm$0.08 & 0.33  \nl
Mar 05 - Mar 16 & 12 & 0.96$\pm$0.08 & $<10^{-38}$ \nl
Apr 07 - Apr 19 & 12 & 1.37$\pm$0.08 & $<10^{-38}$ \nl
Apr 30 - May 15 & 12 & 1.68$\pm$0.09 & $<10^{-38}$ \nl
May 27 - Jun 11 & 14 & 1.61$\pm$0.09 & $<10^{-38}$ \nl
Total      & 55 & 1.29$\pm$0.06 & $<10^{-38}$ \nl

\tableline
\tablevspace{1ex}
\multicolumn{4}{c}{1998}  \nl
\tablevspace{1ex}
\tableline
Feb 28 - Mar 09 & 06 & 1.94$\pm$0.29 & $<10^{-38}$ \nl
Mar 23 - Mar 31 & 03 & 0.00$\pm$0.07 & 1.00         \nl
Apr 19 - Apr 29 & 09 & 0.08$\pm$0.04 & 1.00         \nl
May 20 - May 31 & 10 & 0.09$\pm$0.02 & 0.13        \nl
Jun 16 - Jul 03 & 15 & 0.30$\pm$0.05 & $<10^{-38}$ \nl
Jul 15 - Jul 27 & 06 & 0.12$\pm$0.04 & 0.94        \nl
Total           & 49 & 0.32$\pm$0.04 & $<10^{-38}$ \nl
\tablenotetext{a}{Average rate over the observation period
expressed as a fraction of the rate from the Crab Nebula.}
\tablenotetext{b}{$\chi^2$ probability for constant emission
after accounting for trials.}
\enddata  
\label{table:daily_var}
\end{deluxetable}

\clearpage

\clearpage




\clearpage

\begin{thebibliography}{}

\bibitem[Breslin et al. 1997]{Breslin97}
Breslin, A. C., et al., 1997, IAU Circ., 6592, 1

\bibitem[Buckley et al. 1998]{Buckley98}
Buckley, J. H., et al., 1998, A\&A, 329, 639

\bibitem[Buckley et al. 1996]{Buckley96} 
Buckley, J. H., et al., 1996, \apjl, 472, L9

\bibitem[Catanese et al. 1997]{Catanese97}
Catanese, M., et al., 1997, \apjl, 487, L143

\bibitem[Catanese et al. 1998]{Catanese98}
Catanese, M., et al., 1998, \apj, 501, 616

\bibitem[Cawley et al. 1999]{Cawley99}
Cawley, M. F., et al., 1999, in preparation

\bibitem[Cawley et al. 1991]{Cawley90}
Cawley, M. F., et al., 1990, Exp. Astron., 1, 173

\bibitem[Chadwick et al. 1999]{Chadwick99}
Chadwick, P. M., et al., 1999, \apj, in press

\bibitem[Fegan 1997]{Fegan97}
Fegan, D. 1997, J. Phys. G: Nucl. Part. Phys., 23, 1013

\bibitem[Gaidos et al. 1996]{Gaidos96}
Gaidos, J. A., et al., 1996, Nature, 383, 319

\bibitem[Hillas et al. 1998]{Hillas98}
Hillas, A. M., et al., 1998, \apj, 503, 744

\bibitem[Kataoka et al. 1999]{Kataoka99}
Kataoka, J. et al., 1999, \apj, in press

\bibitem[Kerrick et al. 1995]{Kerrick95}
Kerrick, A. D., et al., 1995, \apj, 452, 588

\bibitem[Krennrich et al. 1997]{Krennrich97}
Krennrich, F., et al., 1997, \apj, 481, 758

\bibitem[Krennrich et al. 1998]{Krennrich98}
Krennrich, F., et al., 1998, \apj, in press

\bibitem[Mattox et al. 1997]{Mattox97}
Mattox, J. et al., 1997, \apj, 476, 692

\bibitem[McEnery et al. 1999]{McEnery99}
McEnery, J. et al., 1999, in preparation

\bibitem[Mukherjee et al. 1997]{Mukherjee97}
Mukherjee, R. et al., 1997, \apj, 490, 116

\bibitem[Protheroe et al. 1998]{Protheroe98}
Protheroe, R. J. et al., 1998, Proc. 25th Int. Cosmic Ray Conf., Durban, 
Highlight Session on Markarian 501, 8, 317

\bibitem[Punch et al. 1992]{Punch92}
Punch, M. et al., 1992, Nature, 358, 477

\bibitem[Press et al. 1988]{Press_recipes}
Press W. H., Flannery, B. P., Teukolsky, S. A., \& Vetterling, W. T, 
1988, Numerical Recipes, The Art of Scientific Computing,
p. 171, Cambridge University Press 

\bibitem[Quinn 1997]{Quinn97}
Quinn, J. 1997, PhD Thesis (unpublished), University College Dublin

\bibitem[Quinn et al. 1996]{Quinn96}
Quinn, J. et al., 1996, \apjl, 456, L83

\bibitem[Samuelson et al. 1998]{Samuelson98}
Samuelson, F. W., et al., 1998, \apj, 501, L17

\bibitem[Schubnell et al. 1996]{Schubnell96} 
Schubnell, M. S., et al., 1996, \apj, 460, 644

\bibitem[Thompson et al. 1995]{Thompson95}
Thompson D. J. et al., 1995, ApJS, 101, 259

\end{thebibliography}
\end{document}